\documentclass[aps,prl,twocolumn,showpacs]{revtex4}

\usepackage{graphicx}
\usepackage{textcomp}

\begin{document}

\title{Observing the Formation of Long-Range Order during Bose-Einstein Condensation}

\author{Stephan Ritter}
\author{Anton \"{O}ttl}
\author{Tobias Donner}
\author{Thomas Bourdel}
\author{Michael K\"{o}hl}
\email[Email: ]{koehl@phys.ethz.ch}
\author{Tilman Esslinger}
\affiliation{Institute for Quantum Electronics, ETH Z{\"u}rich,
8093 Z{\"u}rich, Switzerland}

\date{\today}

\begin{abstract}
We have experimentally investigated the formation of off-diagonal long-range order in a gas of ultracold atoms. A magnetically trapped atomic cloud prepared in a highly nonequilibrium state thermalizes and thereby crosses the Bose-Einstein condensation phase transition. The evolution of phase coherence between different regions of the sample is constantly monitored and information on the spatial first-order correlation function is obtained. We observe the growth of the spatial coherence and the formation of long-range order in real time and compare it to the growth of the atomic density. Moreover, we study the evolution of the momentum distribution during the nonequilibrium formation of the condensate.
\end{abstract}

\maketitle

When a gas of atoms is undergoing Bose-Einstein condensation a macroscopic number of particles start to occupy the same quantum mechanical state---it seems like the randomly colliding atoms are suddenly forced into a lock-step motion. The understanding of this process in which phase coherence spreads over the whole gaseous cloud has intrigued physicists long before Bose-Einstein condensation has been demonstrated \cite{kagan1992,kagan1994}. In particular, the question when the characteristic long-range phase coherence is established is the key point for understanding the condensation process. The trajectory into the state of a Bose-Einstein condensate \cite{kagan1992,kagan1994,svistunov2001,drummond1999} is much more intricate than its equilibrium properties.

The transition to a superfluid or a superconducting quantum phase is a remarkable process in which the properties of the system undergo a fundamental change. The conceptual link between quantum phases in various systems is the off-diagonal long-range order in the density matrix which describes phase correlations over macroscopic distances \cite{penrose1956,yang1962}. The first order spatial correlation function $G^{(1)}(r,r^\prime)=\langle \Psi^\dag(r)\Psi(r^\prime)\rangle$ \cite{naraschewski1999} quantifies the characteristic length scale over which phase correlations exist. Here $\Psi$ ($\Psi^\dag$) denotes the annihilation (creation) operator of the atomic fields. To
experimentally study how long-range order is established in space and time, real time access to this process is required. Yet, due to the short relaxation times and the strong coupling to the environment this seems troublesome in condensed matter samples, such as liquid helium or superconducting materials.

In a trapped atomic Bose gas the situation is distinctly different. It forms an almost closed system with negligible
coupling to the environment and the relaxation time scales are experimentally accessible. Previous experimental studies of the Bose-Einstein condensate formation have focused on the increasing particle density \cite{miesner1998,kohl2002,hugbart2007} and the influence of excitations on the momentum spread in elongated traps \cite{shvarchuck2002,hugbart2007}. The off-diagonal long-range order has been studied only at thermal equilibrium \cite{bloch2000}.

\begin{figure}[b]
\includegraphics[width=\columnwidth]{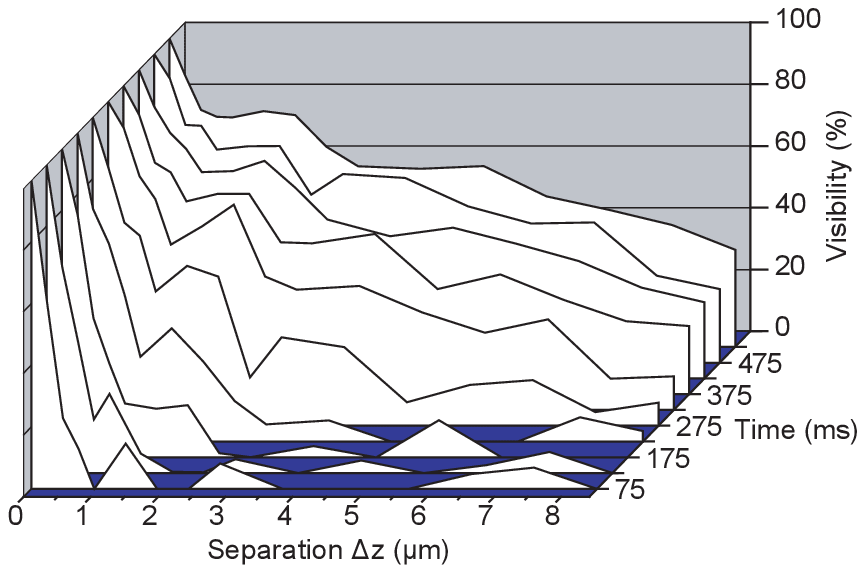}
\caption{Formation of long-range order. Shown is the visibility of a matter wave interference pattern originating from two regions separated by $\Delta z$ inside a trapped cloud. This is a measure of the first-order spatial correlation function of the atomic cloud. By shock cooling the gas is prepared in a highly nonequilibrium state at time $t=0$ and then relaxes towards thermal equilibrium. Initially ($t=75$\,ms) the correlations are short ranged and thermal-like. The onset of Bose-Einstein condensation is marked by the appearance of long-range order extending over all measured distances ($t=275$\,ms). The largest measured distance corresponds to half of the Thomas-Fermi diameter of the final condensate. The data is taken for constant separations versus time. The substructure of the curves reflects the scatter of the initial conditions for each separation. The resulting variance in the onset of condensation is about 30\,ms.}
\label{fig1}
\end{figure}

Here we present an experimental study of the evolution of off-diagonal long-range order during the formation of a
Bose-Einstein condensate out of a nonequilibrium situation. We start from a Bose gas above the phase transition temperature and suddenly quench the gas into a strongly nonequilibrium state \cite{miesner1998,kohl2002,shvarchuck2002,hugbart2007}. Subsequently, the gas can be regarded as a closed system which
evolves into a Bose-Einstein condensed phase. The off-diagonal long-range order is measured by studying the interference pattern of atomic matter waves originating from two different locations in the atom trap \cite{bloch2000}. The visibility of this interference pattern measures the phase coherence between the two regions and its temporal evolution uncovers the formation of long-range order. We continuously monitor the interference pattern during the formation using a single atom counter with high temporal resolution \cite{ottl2005}. Varying the vertical distance $\Delta z$ between the two locations allows us to experimentally map out the evolution of the phase coherence in the trapped Bose gas (see Fig.\,\ref{fig1}).

In the initial nonequilibrium state just after the quench we observe purely thermal-like short-range correlations which decay on a characteristic length scale given by the thermal de Broglie wavelength $\lambda_{dB}\approx 0.4$\,\textmu m. After the first 200\,ms during which little changes are detected, the length scale over which phase correlations exist expands rapidly and the long-range order of a Bose-Einstein condensate is established after a further 100\,ms. Subsequently, a stage of gradual condensate growth towards equilibrium is observed. In the Bose-Einstein condensed phase the system does---apart from its size---not exhibit a characteristic length scale \cite{landau1980}.

In the theoretical investigation of the nonequilibrium formation of a Bose-Einstein condensate one distinguishes qualitatively between a kinetic evolution and a coherent evolution \cite{kagan1992,gardiner1998,bijlsma2000,svistunov2001}. During the kinetic stage of the evolution, the occupation numbers of the low-energetic states grow. It is governed by elastic collisions and the characteristic time scale is set by the collision time $\tau_\mathrm{col} = (n \sigma v_T)^{-1}$ with $n$ being the peak density of the gas, $\sigma$ the elastic collision cross section, and $v_T$ the average thermal velocity. The coherent stage---which plays a significant role only in the regime of very large density or scattering length---describes the merging of quasicondensates, patched regions of locally constant phase, into a full Bose-Einstein condensate. The presence of quasicondensates could be revealed by the momentum distribution of the gas below the critical temperature.

Our experimental setup and the procedure for Bose-Einstein condensation have been described previously \cite{ottl2006}. For the experiments reported in this Letter, we start by preparing a thermal cloud of atoms in the $|F = 1, m_F = -1 \rangle$ hyperfine ground state of $^{87}$Rb in a harmonic magnetic trapping potential. The trapping frequencies of the magnetic trap are $(\omega_x,\omega_y,\omega_z)= 2 \pi \times (39,7,29)$\,Hz, where $z$ denotes the vertical axis. The temperature of the atom cloud $T = 240$\,nK is slightly above the transition temperature $T_c = 215$\,nK for Bose-Einstein condensation for the given number of atoms $1.3\times 10^7$. This results in a collision time of $\tau_\mathrm{col}\approx 30\,$ms. Temperature and atom number are measured by absorption imaging. The atom cloud is then brought into a strong nonequilibrium situation by rapidly lowering the trap depth to 620\,nK and removing the high-energy tail of the Maxwell-Boltzmann distribution. Within the 100\,ms of this ``shock cooling'' we remove $30\%$ of the atoms. Subsequently, the cloud relaxes from its highly nonequilibrium state and within a few hundred milliseconds $3\%$ of the atoms form a Bose-Einstein condensate. During the relaxation process particle number and total energy are conserved with minimal disturbance due to the detection process. Moreover, we do not observe oscillations of the condensate with respect to the detector as a consequence of the shock cooling.

We detect the evolution of both the density and the long-range order of the cloud simultaneously and in real time using radio frequency output coupling \cite{bloch1999} and single atom counting \cite{ottl2005,bourdel2006}. For output coupling we apply a weak monochromatic microwave frequency field which spin-flips atoms from the magnetically trapped state into the untrapped state $|F = 2, m_F = 0 \rangle$. The untrapped atoms form a downwards propagating atomic beam. The output coupling region is defined by a surface of constant magnetic field and can be approximated by a horizontal plane within the atomic cloud \cite{bloch1999}. Applying two microwave fields with different frequencies realizes output coupling from two vertically separated surfaces of constant magnetic field \cite{bloch2000} which are chosen symmetric about the center of the cloud. The two overlapping atomic beams interfere with each other. The mean flux reflects the local density of atoms at the position of output coupling within the transverse momentum interval measured by our detector. The visibility of the interference pattern reflects the phase coherence. The interference pattern is detected in time with single atom resolution using an ultrahigh finesse optical cavity, located 36\,mm below the magnetically trapped atoms \cite{ottl2005}. The presence of an atom inside the cavity leads to a decrease in the transmission of a probe laser beam resonant with the empty cavity \cite{mabuchi1996}. Our system permits a precise measurement of the time-dependent atom flux.

\begin{figure}
\includegraphics[width=0.95\columnwidth]{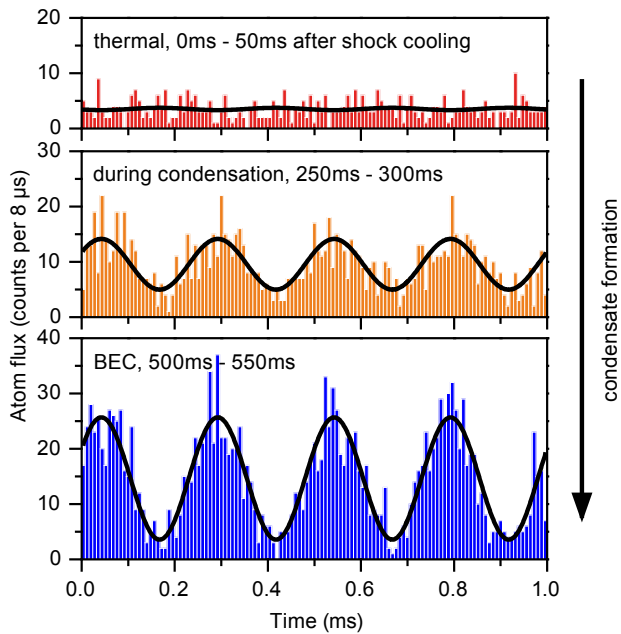}
\caption{Buildup of density and off-diagonal long-range order during condensate formation. We plot the histogram of the atom arrival times modulo 1\,ms for $\nu=4$\,kHz. The data are summed over 18 repetitions of the experiment. The black lines are fits to the data to extract the mean atom flux $A$ and the visibility $V$ of the interference pattern (see text).}
\label{fig2}
\end{figure}

The mean atom flux and the visibility of the atomic interference pattern are determined in time bins of 50\,ms length. In Fig.\,\ref{fig2} three different situations of the condensate formation are shown. For the cloud immediately after shock cooling [Fig.\,\ref{fig2}(a)], the visibility is zero and the mean atom flux is low. As the condensation process develops [Fig.\,\ref{fig2}(b)], the atom flux increases and interference arises. Both continue to grow up to a final value [Fig.\,\ref{fig2}(c)] determined by the condensate fraction and detector function. The black lines are a fit of $f(t) = A\left[1 + V \sin\left( 2 \pi \nu t +\phi \right) \right]$ to the data. The frequency $\nu$ is the difference of the two microwave frequencies used for output coupling. The phase $\phi$ is determined by the relative phase of the two microwave fields which is locked to the experimental cycle. The output coupling of atoms with a rate $\approx 5 \times 10^4$\,atoms per second is essentially not perturbing the density dynamics of the condensate formation. In particular, it has no detectable influence on the final condensate fraction.

The observation of long-range order even at a very small condensate fraction poses a severe technical challenge. Our
detection scheme has specific properties which render the present measurement possible. The output coupling probability for a condensed atom is a factor of 4 larger than for a thermal atom due to their different density distributions for our equilibrium parameters. Moreover, we observe unequal detection efficiencies for condensed and noncondensed atoms which we estimate to be $23\%$ and $1\%$, respectively. They are mainly determined by the size of the atomic beam exceeding the active area of the detector. For thermal atoms the transverse velocity spread is larger which reduces the probability for an atom to be detected in the cavity \cite{ottl2006}. Because of these effects the visibility of the interference pattern greatly exceeds the condensate fraction and reaches a maximum of $70\%$ for $\Delta z > \lambda_{dB}$. We observe a reduction of the visibility towards large values of $\Delta z$ (Fig.\,\ref{fig1}). This is influenced by the following effects: Output coupling symmetrically off center from the Thomas-Fermi profile leads to a reduction of $7\%$ at $\Delta z=9$\,\textmu m. Furthermore, the spatial modes of the two interfering atom laser beams depend on their divergence and thus on the position of output coupling \cite{lecoq2001}. We calculate this effect to be $10\%$ for the largest $\Delta z$. An additional contribution is due to the geometry of our trapping potential resulting in a tilt of the long axis of the condensate by $4^\circ$ with respect to the horizontal. Moreover, the residual uncertainty in determining the arrival times of the atoms (4\,\textmu s) limits the attainable contrast for the largest slit separations in our experiment to $94\%$.

\begin{figure}
\includegraphics[width=0.95\columnwidth]{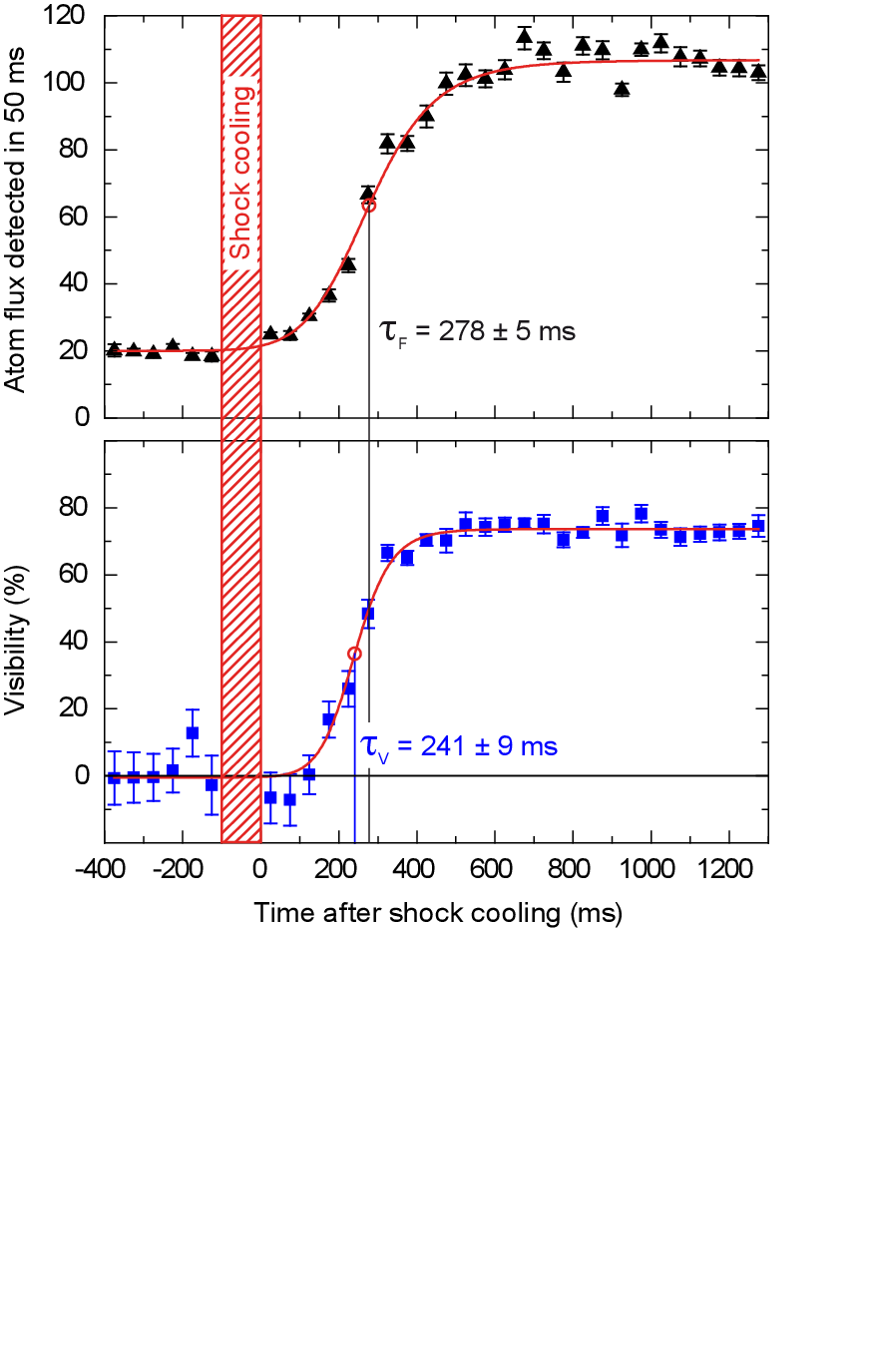}
\caption{Growth curves for atom flux and visibility of the interference pattern during the formation of a Bose-Einstein condensate. The separation $\Delta z$ between the two output coupling regions is 1.9\,\textmu m. Before the shock cooling stage there is a low flux of thermal atoms and no interference. The growth in atom flux is significantly slower than the growth of the interference pattern. The data are averaged over 20 repetitions and the error bars indicate the statistical error of these measurements which are within a factor of 2 of the theoretical shot-noise limit.}
\label{fig3}
\end{figure}

In Fig.\,\ref{fig3} we show the growth of the density reflected in the mean atom flux and of the visibility for a given separation of the output coupling regions. This corresponds to a section of constant $\Delta z=1.9$\,\textmu m in Fig.\,\ref{fig1}. We have analyzed both the delay of the formation with respect to the shock cooling stage and the speed of the formation for density and off-diagonal long-range order. We fit a function $g(t) =A_2 + (A_1-A_2)/[1+(t/\tau)^p]$ to the data to quantify the growth, where $\tau$ denotes the time after which $50\%$ of the increase in flux or visibility are reached.

The duration of the condensate formation can be quantified by the time needed for an increase of both the flux and the visibility from $10\%$ to $90\%$ of the total increase. We find the duration to be approximately independent of the separation $\Delta z$ of the two output coupling regions. For the flux this time is $421 \pm 47$\,ms whereas the visibility of the interference pattern grows faster in a time of $267 \pm 48$\,ms. From the data we find that the time $\tau$ after which $50\%$ of the total atom flux is reached is almost constant across the investigated inner region of the trapped cloud, whereas for the visibility it increases with larger slit separation. The coherent region grows with a velocity of $\approx 0.1$\,mm/s, approximately a factor of 5 slower than the speed of sound at the peak density of the
thermal cloud. The speed of sound imposes a natural speed limit for the expansion of the coherent region. While the onset of the growth (defined by the $10\,\%$ points) starts $39 \pm 54$\,ms earlier for the flux than for the coherence, the visibility saturates ($90\,\%$ points) about $116 \pm 34$\,ms earlier than the density. This applies to all separations $\Delta z$ larger than the de Broglie wavelength.

Moreover we find that $\tau$ decreases with increasing size of the final condensate. To obtain larger condensates in equilibrium, the initial particle number or temperature or the fraction of atoms removed during shock cooling was changed. As the limiting cases we find for a final condensate number of $2.5\times10^5$ atoms $\tau=(282,320)$\,ms and for $8.9\times10^5$ atoms $\tau=(57,114)$\,ms, where the first number in the bracket refers to the visibility and the latter to the flux. The numbers are averaged over all slit separations. This decrease of the formation time for larger condensates is proportional to the increase of the elastic collision rate due to the higher final density.

\begin{figure}
\includegraphics[width=0.95\columnwidth]{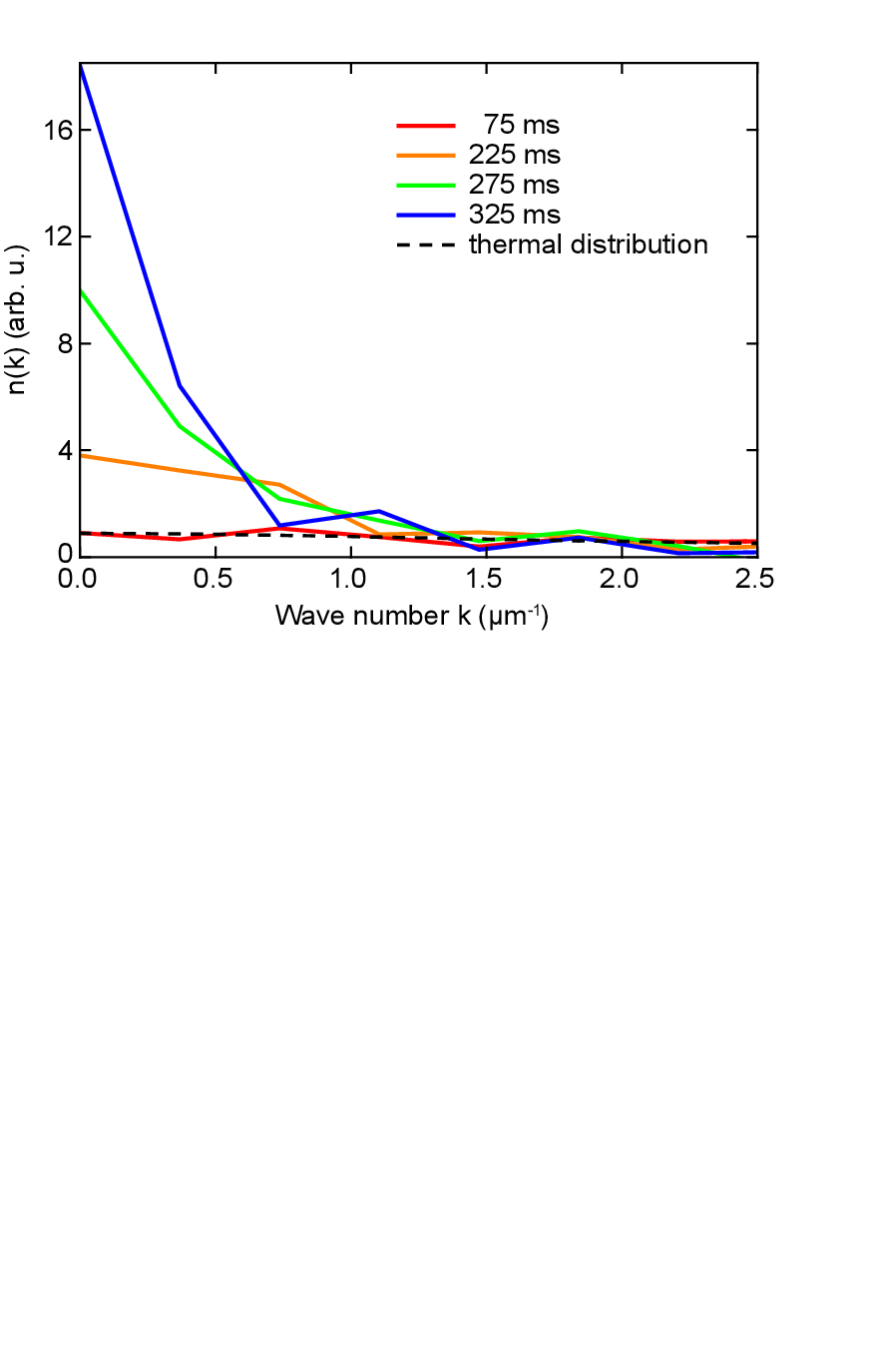}
\caption{Temporal evolution of the momentum distribution of the detected atoms. The distribution is obtained by Fourier transformation of the visibility data. The thermal distribution is calculated for the parameters before shock cooling including a normalization factor.}
\label{fig4}
\end{figure}

It is instructive to study the evolution of the vertical momentum distribution $n(k)$ during the thermalization process. We infer the distribution $n(k)$ of the detected atoms from the Fourier transform of the visibility data of Fig.\,\ref{fig1} multiplied by the average flux \cite{gerbier2003}. The result is shown in Fig.\,\ref{fig4}. For the initial state directly after the quench, the momentum distribution is well approximated by a thermal distribution (dashed line). The thermalization process after the shock cooling leads to the formation and the growth of a low-momentum peak which in equilibrium corresponds to the condensate. For our parameters in equilibrium  we expect a true condensate without strong residual phase fluctuations \cite{petrov2001a}. In Fig.\,\ref{fig4}, we focus on the beginning of the condensate formation. After 225\,ms of evolution, we observe a broader low-momentum peak (width in $k\sim 0.8\,\textrm{\textmu m}^{-1}$) than for the final distribution. For evolution times of 325\,ms and larger, the width of the low-momentum peak is limited by our measurement resolution of $2 \pi/(16\,\textrm{\textmu m)}$, which takes into account that the correlation function is symmetric with respect to $\Delta z=0$.

In summary, we have experimentally studied the dynamics of Bose-Einstein condensation and more specifically the formation of long-range order using an interference measurement. We have quantitatively compared the evolution of the flux and of the visibility of the interference pattern. Finally, we have presented our data in terms of the atom momentum distribution.

We would like to acknowledge helpful discussions with F.T. Arecchi, G. Blatter, F. Brennecke, Y. Kagan, and A. Kuklov. T.B. acknowledges funding by an EU Marie Curie Action under contract MEIF-CT-2005-023612.

\end{document}